\documentclass[aps, prb, numerical, superscriptaddress, preprint]{revtex4-1}

\usepackage{amsmath}  
\usepackage{amsfonts} 
\usepackage{graphicx} 

\begin{document}
\bibliographystyle{ajp} 

\title{Quantifying biomolecular diffusion with a ``spherical cow'' model}

 \author{Frederico Campos Freitas}
 \affiliation{Laborat\'orio de Biof\'isica Te\'orica, Departamento de F\'isica,
Instituto de Ci\^encias Exatas, Naturais e Educa\c{c}\~ao, Universidade Federal do Tri\^angulo Mineiro, Uberaba, MG, Brazil.}

\author{Sandra Byju}
\affiliation{Department of Physics and Center for Theoretical Biological Physics, Northeastern University, 360 Huntington Avenue, Boston, MA 02115}
\author{Asem Hassan}
\affiliation{Department of Physics and Center for Theoretical Biological Physics, Northeastern University, 360 Huntington Avenue, Boston, MA 02115}
\author{Ronaldo Junio de Oliveira}
\affiliation{Laborat\'orio de Biof\'isica Te\'orica, Departamento de F\'isica,
Instituto de Ci\^encias Exatas, Naturais e Educa\c{c}\~ao, Universidade Federal do Tri\^angulo Mineiro, Uberaba, MG, Brazil.}

\author{Paul C. Whitford}
\email{To whom correspondence should be addressed: p.whitford@northeastern.edu}
\affiliation{Department of Physics and Center for Theoretical Biological Physics, Northeastern University, 360 Huntington Avenue, Boston, MA 02115}

\date{\today}

\begin{abstract}
The dynamics of biological polymers, including proteins, RNA, and DNA, occur in very high-dimensional spaces. Many naturally-occurring polymers can navigate a vast phase space and rapidly find their lowest free energy (folded) state. Thus, although the search process is stochastic, it is not  completely random. Instead, it is best described in terms of diffusion along a downhill free energy landscape. In this context, there have been many efforts to use simplified representations of the energetics, for which the potential energy is chosen to be a relatively smooth function with a global minima that corresponds to the folded state. 
That is, instead of  including every type of physical interaction, the broad characteristics of the landscape are encoded in approximate energy functions. We describe a particular class of models, called structure-based models, that can be used to explore the diffusive properties of biomolecular folding and conformational rearrangements. These energy functions may be regarded as  the ``spherical cow''  for modeling molecular biophysics. We discuss the physical principles underlying these models and  provide an entry-level tutorial, which may be adapted for use in  curricula for physics and non-physics majors.
\end{abstract}

\maketitle

\section{\label{sec:intro}Introduction}

When studying a complex system,  physicists will typically begin by proposing  a highly simplified model that includes a few relevant properties of the system. The broad utilization of this strategy inspired the well-known joke in the physics community regarding a ``spherical cow,'' with several examples of this approach immortalized by a book entitled by the joke.~\cite{Harte1988} That is, when studying a cow, a physicist's first approximation is to represent the cow    by a sphere of uniform mass and charge density. Starting with this spherical cow, physicists will then  investigate the properties of the simplified system before considering additional details. By iteratively introducing new features, complex physical systems can be understood at ever-increasing levels of detail. In  contrast to this approach, traditional biological studies aim to provide broad characterizations (e.g., structures and rates) of detailed systems (e.g., molecules in a cell). Thus, at first glance, it may not be obvious how physicists can effectively apply the spherical cow philosophy to biology.

In the following, we will discuss a spherical cow approach to studying molecular biophysics.  Specifically, we will explain the ideas behind a class of potential energy functions called structure-based models.~\cite{Clementi:2000p3234,whitford2009all,SMOG2}  These models exploit the   phenomenological features of biomolecules   to provide a  simplified version of the energetics. To understand the value of these models, it is necessary to recognize that molecular biology techniques can  provide only atomic-resolution descriptions of long-lived stable structures of biomolecules. Accordingly,  these configurations must correspond to deep (at least several $k_BT$) free energy minima. Inspired by this simple observation, structure-based models explicitly define  experimental configurations to be stable.  That is, the baseline versions of these models do not aim to identify the factors that impart stability.  Rather, interactions formed in the native (ground state) configurations are defined to be stabilizing, and all other interactions are treated as repulsive, which ensures that the spatial arrangements are preserved. Given the crude character of the models, it may be surprising that these simplified representations have been able to model a broad range of biomolecular processes, ranging from protein folding~\cite{Clementi:2000p3234,Cheung:2003p3381,Chavez:2004p3380,Yang:2004p3361} to the dynamics of protein synthesis by the ribosome.~\cite{Noel:2016bv,LeviMethods,Levi2020}

We first provide a brief introduction to molecular biology for physics students,   followed by a  description of simulation techniques and structure-based models. We additionally discuss example calculations that can be adopted and integrated in advanced undergraduate or graduate-level physics courses.  Our intent    is to provide  students (and instructors) with a basic understanding of the biological context and physical principles. To facilitate the adoption of this material, we provide a repository with step-by-step instructions on how to apply the  models in simulations.

\section{Molecular Biology: A primer for physics students}

Because  most physics   students have limited exposure to biological systems, we  first provide some  biochemical and physical-chemical background that is necessary  for  understanding the physical principles that govern biology. We will focus on biological polymers, including proteins and nucleic acids. Due to the intimate relationship of structure and function, we will discuss both chemical composition and empirically-determined structural properties. 
Although our discussion can be found in  standard biochemistry texts,  our overview  allows for a more focused entry into the simulation of biopolymers.

\subsection{Protein structure}
\begin{figure}[htbp!]
	\centering
	\includegraphics[width=3.4in]{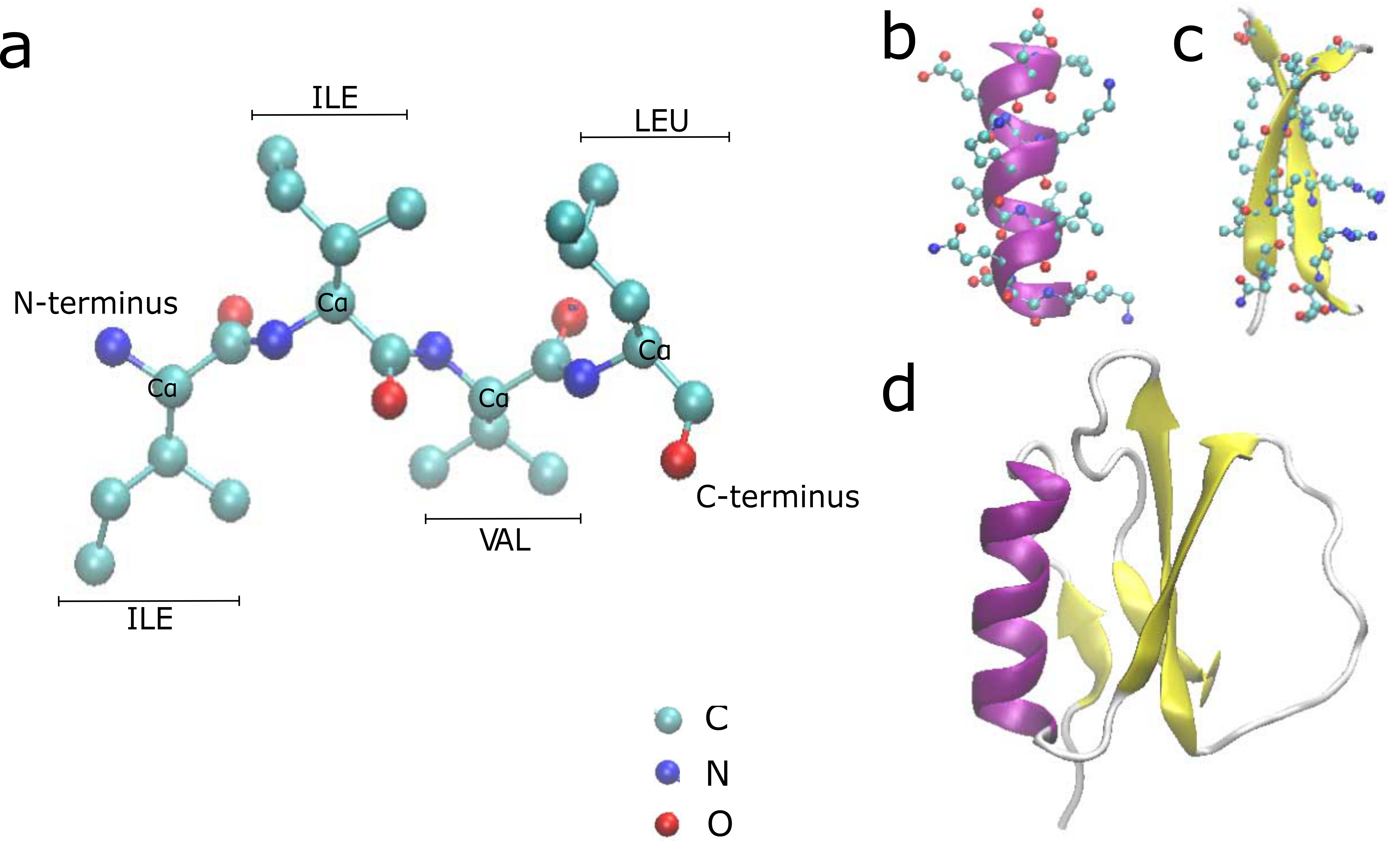}
	\caption{(Color online) Protein Structure. (a) A polypeptide composed of four amino acid residues; the N- and C-terminal ends are marked. Shown is a ball and stick representation of all non-hydrogen atoms (color coded), with bonds shown as thin cylinders. (b) An $\alpha$ helix of  protein chymotrypsin inhibito (CI2) with helical ribbons representing the polypeptide backbone and side chains shown explicitly using a ball and stick representation. (c) A $\beta$ sheet of CI2 with aligned arrows representing the polypeptide backbone and side chains shown in a ball and stick representation. (d) Tertiary structure of CI2. Cartoon representation of CI2 in its folded conformation, with $\alpha$ helix (purple) and $\beta$ sheets (yellow) highlighted. }
	\label{protein}
\end{figure}

Proteins have many roles in the cell,  including providing structural integrity, executing chemical reactions, signaling, and regulating gene expression. A protein is a polymer that is formed by a sequence of amino acid residues. Each amino acid  or residue [see Fig.~\ref{protein}(a)] is composed of a common amino group (NH$_2$), carbon ($C_\alpha$), and carboxyl group (CO$_2$), while the  ``side chain'' (usually denoted by the letter R)  differs for each type of residue.  There are 20 naturally occurring amino acids,  each  defined by the composition of the R group.  Each amino acid is linked to the next residue by the formation of a peptide bond, such that a single protein chain is  called a ``polypeptide.'' When describing protein structures, the N-terminal end is considered the ``beginning'' of the chain, and the end of the chain is the C-terminal tail. If we exclude the side chain atoms,  we can define the ``backbone'' of the chain    by the repeating set of common atoms. Because the backbone atoms are common to each residue,  amino acids are  generally classified based on the side chain (R) composition, where they can be  acidic, basic, uncharged polar, and non-polar (hydrophobic). 

There are several dominant classical forces that  describe the energetics of proteins. Along with the covalently-linked backbone atoms, there are weaker non-covalent interactions that can be formed between all atoms in the chain. The four types of non-covalent interactions that are relevant in biomolecules are electrostatic, hydrogen bond, van der Waals, and hydrophobic interactions. Charged species interact via long-range Coulomb interactions. Hydrogen bonds are directional dipole-dipole interactions that can be formed between highly electronegative atoms (donors), such as N, O or F, with electronegative atoms (acceptor atoms). The van der Waals interaction accounts for the excluded volume due to the exclusion principle, as well as a net attractive force due to instantaneous dipole-dipole interactions. Hydrophobic interactions describe the way by which non-polar hydrophobic (water-repelling) residues favor aggregation to minimize exposure to the polar solvent environment.

To systematically describe a protein, it is necessary to decompose its structure  into multiple tiers. At the most basic level, the primary structure is used to define the sequence of amino acids that are present in a single chain as the four-amino acid polypeptide in Fig.~\ref{protein}(a). Local structure formation in short segments (typically 10--20 residues) is called the secondary structure, where the two major structural motifs are the $\alpha$ helix and $\beta$ sheet [see Figs.~\ref{protein}(b) and~\ref{protein}(c)]. In an $\alpha$ helix, the polypeptide twists to form a right-handed helical structure that is stabilized by hydrogen bonds formed along the protein backbone. To aid the inspection of the structure,  graphical software will typically display these regions as helical ribbons [Fig.~\ref{protein}(b)]. 
The second common structural motif is the $\beta$ sheet, which is formed when two or more spatially-adjacent segments of the polypeptide chain align in a parallel or anti-parallel arrangement. These elements are often shown as aligned arrows [Fig.~\ref{protein}(c)]. At a higher level of organization is the tertiary structure of a protein, which typically  involves spatial organization of numerous secondary structure elements [Fig.~\ref{protein}(d)].  When a portion of the chain assembles into an autonomous unit, these elements are often called ``domains.'' The final level of organization is the quarternary structure, which refers to the assembly and association of multiple peptide chains.

\subsection{Structural properties of nucleic acids (RNA and DNA)}
\begin{figure}[htbp!]
	\centering
	\includegraphics[width=3.4in]{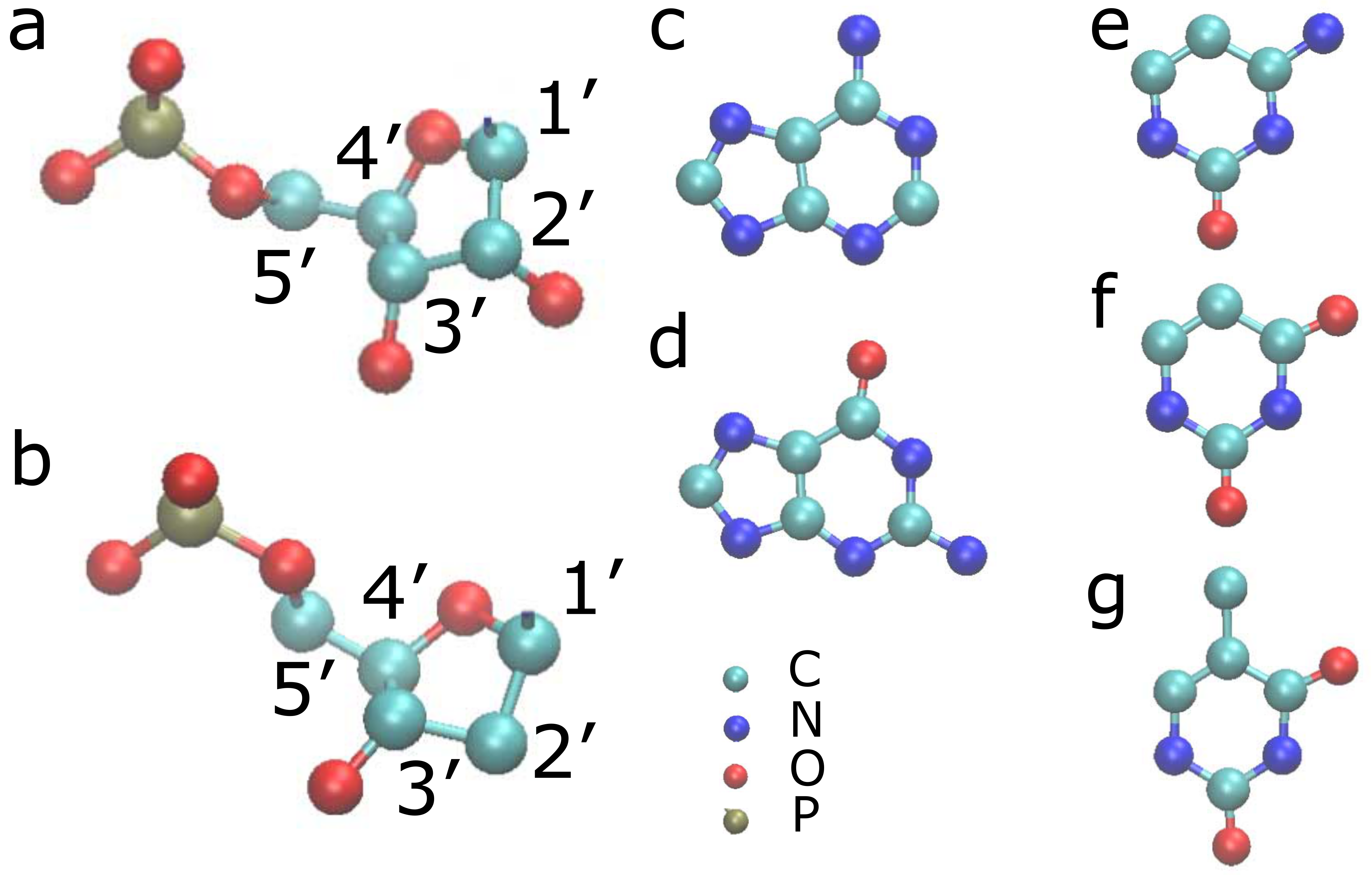}
	\caption{(Color online) Nucleic acids. Each nucleotide is composed of a pentose sugar with phosphate group attached to 5' carbon and nucleobase attached to the 1' carbon. (a) RNA nucleoside with ribose sugar and a phosphate group connected to the 5' carbon. (b) DNA nucleoside with deoxyribose sugar and a phosphate group connected to the 5' carbon. Purine nucleobases are  Adenine (c) and  Guanine (d). Pyrimidine nucleobases are Cytosine (e), Uracil (f) and Thymine (g); Uracil is unique to RNA as Thymine is to DNA.}
	\label{rna-dna}
\end{figure}

The second class of biomolecules that we will discuss are nucleic acids, which are formed by a string of nucleotide monomers. A nucleotide is composed of a backbone pentose sugar, where the carbon atom in the 5' position is linked to a phosphate group, and a nitrogenous base is covalently bonded to the 1'-carbon atom through a N-glycosidic linkage [Figs.~\ref{rna-dna}(a) and~\ref{rna-dna}(b)]. Nucleic acid carbons belonging to sugar rings have the prime symbol added to their numbers to differentiate them from nucleobase ring carbons. In contrast to proteins, which carry their $+1$ or $-1$ charge on the side chain, the nucleic acid backbone has a negative charge that arises from the phosphate group (PO$_4^-$). As for  proteins, the sequence of nucleic acid residues defines the polymer. On each residue, there is a nitrogenous base, which is either a purine (Adenine, Guanine [Figs.~\ref{rna-dna}(c) and~\ref{rna-dna}(d)]), or a pyrimidine (Cytosine, Uracil, Thymine [Figs.~\ref{rna-dna}(e),~\ref{rna-dna}(f), and~\ref{rna-dna}(g)]). The two major classes of nucleic acids (RNA and DNA) are classified by the presence/absence of a single oxygen atom on the 2' carbon in the backbone of each residue [Figs.~\ref{rna-dna}(a) and~\ref{rna-dna}(b)].

DNA typically forms the well-known double-stranded helix, where hydrogen bonds are formed between complimentary bases in the two chains. This stabilizing energy imparted by Watson-Crick base pairing is in addition to that associated with the stacking of adjacent bases (hydrophobic and van der Waals forces). The most common DNA structure is the right-handed double helix (B form).

In contrast to DNA,  RNA molecules    adopt a wide range of stable conformations. With this extended versatility, they can contribute to functional conformational dynamics in the cell by serving as biomolecular machines (e.g., the ribosome) or performing enzyme activity (e.g., ribozymes). Although all RNA molecules have similar components, their functional roles have led to the introduction of many different names.  For example, RNA associated with gene expression is often called mRNA, and RNA present in the ribosome is called rRNA. Regardless, RNA can exist in isolation (i.e., individual chains), where hydrogen bonds and Coulomb forces stabilize a range of structural motifs, including the RNA double helix (A form), hairpin loops, internal loops, bulges and junctions.  Together, these structural elements are referred to as the secondary structure of an RNA molecule.  These secondary structure elements  arrange to form functional tertiary conformations. In addition to RNA-RNA interactions, the large negative charge of RNA chains leads to a strong dependence of RNA structure on metal ions, including Mg$^{2+}$ and K$^{+}$.

\subsection{Protein synthesis, folding and assembly}
DNA, RNA, and proteins are the key biomolecular components that define  molecular biology. That is, genetic information is coded in DNA, which is then transcribed to  mRNA sequences.  Although the transcription of mature mRNA in prokaryotes (single-cell organisms) is performed by RNA polymerase, in eukaryotes (multi-cellular organisms), the mRNA can be further modified through a range of processes, such as splicing. In both cases, the mature mRNA sequence is read and translated into a protein sequence by the ribosome. After the ribosome produces a new protein, the protein must then fold   to carry out any given biological purpose.

The most common way to begin to think about protein folding is to begin with what is known as Levinthal's paradox.~\cite{levinthal1969fold} According to this ``paradox,'' a random search to find the folded conformation would require the age of the universe.  For example, for a polypeptide of 100 amino acid residues, with each residue having two allowed conformations, there would be a total of $2^{100}$ possible conformations. If the random sampling of conformations occurred every picosecond, it would require roughly $10^{10}$ years for a single protein to fold. However, protein folding occurs many orders of magnitude faster, generally between microseconds and seconds. 

To reconcile the apparent paradox, it was recognized that the energy landscapes of proteins must not be random or flat. Instead, the energy landscapes may be thought of as being funnel-shaped,~\cite{Leopold:1992p11568} where the principle of minimal frustration~\cite{Bryngelson:1987p11736,BRYNGELSON:1989p1699,BRYNGELSON:1990p6116,BRYNGELSON:1995p1708} indicates that there is a lack of large-scale energetic traps. In this framework, protein folding may be described as a diffusive process, where the protein moves along the landscape in the direction of the global minimum. Due to the presence of a large energy gap between the native and unfolded conformations relative to the scale of the energetic roughness, the dynamics in these funneled landscapes  yields time scales that are consistent with the dynamics in the cell.

\section{Theoretical Models and Computational Methods}

Although experimental molecular biology techniques can determine the atomic structures of stable structures, describing their dynamics requires an understanding of  energetics. From a theoretical/computational perspective, we need   to specify how atoms within a biomolecule interact, typically by constructing a suitable potential energy function. Once a potential energy function (force field) is defined, we  use numerical techniques to evaluate the associated kinetic and thermodynamic properties of the system. The two most widely used computational techniques are Monte Carlo and Molecular Dynamics (MD) simulations. For a  introduction, see Ref.~\onlinecite{sims}. We will focus on MD techniques, because they are more widely used to study the kinetics and equilibrium distributions in proteins. We first describe the physical principles behind MD simulations,   followed by a discussion of  the stypes of force fields that may be applied in simulations of biomolecules.

\subsection{Molecular Dynamics simulations}
Molecular dynamics simulations involve numerically integrating Newton's equations of motion, using a discrete time step $\tau$ (generally femtosecond scale). At each step, the molecular forces are calculated, the coordinates and velocities are incremented in time, and the process is then repeated.~\cite{sims} In its simplest form, we can use the positions and velocities at time $t$, along with the force ($\mathbf{{F}}=-\boldsymbol{\nabla} U$) to determine the positions and velocities at time $t+\tau$ by the relations:
\begin{align}
\mathbf{X}(t+\tau)&=\mathbf{X}(t)+\mathbf{V}(t)\tau\\
\mathbf{V}(t+\tau)&=\mathbf{{V}}(t)+\frac{\mathbf{{F}}(t)}{m}\tau.
\end{align}
Although higher-order approximations, such as the Verlet algorithm,~\cite{PhysRev.159.98} are typically applied in research settings, all MD simulations share this  approach for determining the time sequence of configurations. By using this simple approach, a trajectory of the motion can be obtained to probe functional conformational transitions,~\cite{levi2019dissecting} ligand binding,~\cite{Mobley:2007p1741} subunit association,~\cite{kim:2014fv} and folding transitions.~\cite{EastwoodWolynes2001,LindorffLarsen:2011p11712} We will focus on protein folding/unfolding transitions to illustrate the methods and physical analyses that are available.

To integrate Newton's equations of motion, the potential energy function must be defined. The potential energy function contains terms that specify the nature of the interactions between bonded and non-bonded atoms. Bonded interactions are typically associated with pairs of atoms, triplets that form angles, or quartets that form dihedral angles. These bonded terms define the covalent bond geometry of the biomolecule, and  approximate the vibrational properties that arise from covalent interactions. 

Interactions between non-covalently-bonded atoms can be divided into several broad classes. Steric repulsion arises from atomic exclusion, where two atoms are not allowed to overlap their electronic densities. In classical MD simulations, this effect can be described by many possible functional forms, although a $1/r^{12}$ relation is most commonly applied. Another major contributor is electrostatic interactions. Because classical MD simulations do not explicitly describe electrons, partial charges are typically assigned to each atom, and the precise values are intended to reflect the associated (average) electronic densities. These charges  interact via Coulomb potentials, screened-Coulomb potentials, or other implicit-solvent representations. In addition to electrostatics, dispersion forces can also lead to attraction between atoms. These effects are commonly approximated by a $1/r^{6}$ dependence. As we have described, hydrogen bonds can also occur, where proton-mediated interactions are formed between two highly electronegative atoms. Finally, base-stacking interactions are generally attractive in RNA and DNA and  provide a critical contribution to overall molecular stability.

A particularly important factor that influences the structure of biomolecules  is the nature of solvent (water) interactions. Sometimes, structural water molecules can bind to proteins and form stabilizing hydrogen bonds with acceptor groups. In addition, the solvent can mediate hydrophobic interactions, which are entropic in nature. The hydrophobic effect is due to hydrophobic residues being sequestered from the solvent, which leads to an increase of the configurational entropy of the solvent. We will perform simulations with an implicit-solvent model for which the impact of the solvent is included by an effective representation.

After defining the functional form of different terms in the potential energy function, parameters  have to be specified. The parameters can be determined using quantum mechanical calculations,  semi-empirical comparisons,  effective  parameterization strategies, or phenomenologically-based arguments. In the latter cases, the energetic terms represent effective interactions between atoms inside the biomolecule. These effective interactions typically account for solvent effects when the solvent is not explicitly modeled, and any other energetic interactions that are not represented explicitly, which can include hydrogen bonds, salt bridges, and structural water molecules.
 
 \subsection{Coupling to a heat bath}

Biomolecules are constantly bombarded   by collisions with water and other molecules inside the cell. To describe the exchange of energy between the molecule of interest and the local environment, it is common for MD simulations to couple the dynamics to an external constant temperature heat reservoir. 
If the number of particles and volume are also held constant, then the canonical NVT ensemble  is described. One way to account for coupling to a heat bath is to apply Langevin dynamics.  In these applications, we integrate Newton's equations of motion, and the effect of the solvent is taken into account by introducing a drag term and a random force term:
\begin{equation}
m\mathbf{\ddot{X}}=-\boldsymbol{\nabla} U-\gamma \mathbf{\dot{X}} + \sqrt{2\gamma k_B T}\mathbf{R}(t),
\end{equation}
where $\mathbf{X}$ is the position of an atom (out of $N$ atoms), $U$ is the potential energy function, and $\gamma$ is an effective drag coefficient. The second term represents dissipative momentum exchange of the protein with the solvent molecules, and the third term represents the random force imparted by collisions with the solvent atoms. This random force is represented as Gaussian white noise with zero mean, and the distribution of values is defined according to the fluctuation-dissipation theorem.~\cite{Kubo_1966}

\subsection{Structure-based models: The ``spherical cow''}
As discussed in Sec.~\ref{sec:intro}, physicists usually  begin their study of a complex system by proposing a  simplified model (think of how many times you have reduced a system to a simple harmonic oscillator).  In the context of understanding the physics of molecular biology, an extremely simplified version of a biomolecule is a structure-based model, which we now describe.

We will employ an all-atom structure-based (SMOG) model~\cite{whitford2009all,SMOG2}  to demonstrate how simple force fields can be used in conjunction with MD simulations to study diffusive aspects of protein folding. In this model, all non-hydrogen atoms are represented as beads of unit mass. The parameters of the force field are  defined to stabilize a known biomolecular structure, which is usually obtained from experiment. The force field is intended to represent the effective energetics of the system, after averaging over electrostatics, van der Waals,  and solvent effects. With this approach, the potential energy landscape is funneled toward the native conformation. Accordingly, at sufficiently low temperatures, the free energy landscape will also exhibit funnel-like characteristics. In this representation, the absence of non-native attractive interactions is in accordance with the the principle of minimal frustration.~\cite{BRYNGELSON:1990p6116,Bryngelson:1987p11736,BRYNGELSON:1989p1699} That is, the principle of minimal frustration implies a smooth energy landscape, in which native interactions provide the dominant contribution to the energetics. Although the energy landscape is minimally frustrated, free energy barriers can still occur as a result of steric or entropic factors.  

In this model the potential energy function is given by
\begin{eqnarray}
U &=&\sum_{\rm bonds} \frac{\epsilon_{r}}{2} (r-r_0)^2 + \sum_{\rm  angles} \frac{\epsilon_{\theta}}{2} (\theta-\theta_0)^2 \nonumber \\ 
&&{}+ \sum_{\rm  impropers} \frac{\epsilon_{\chi}}{2} (\chi-\chi_0)^2 +\sum_{\rm backbone-dihedrals} \epsilon_{\rm bb}  F(\phi-\phi_0) \nonumber \\
&&{}+ \sum_{\rm  sidechain-dihedrals} \epsilon_{\rm sc}  F(\phi-\phi_0) \nonumber \\
&&{}+ \sum_{\rm  contacts} \epsilon_c \left[\left( \frac{\sigma_{ij}}{r_{ij}} \right)^{12}-2\left( \frac{\sigma_{ij}}{r_{ij}} \right)^6\right] \nonumber \\
&&{}+ \sum_{\rm  non-contacts} \epsilon_{\rm nc} \left(\frac{\sigma_{\rm nc}}{r_{\rm nc}}\right)^{12}, \label{eq.:SMOG-potential}
\end{eqnarray}
where 
\begin{equation} F(\phi)=\left[1-\cos \phi\right]+\frac{1}{2}  \left[1-\cos3 \phi \right]. \end{equation}

The bond, angle, and improper dihedral terms define the covalent bond structure of the protein. The values of the parameters $r_0$ and $\theta_0$ are taken   from the experimentally-obtained structure. Non-planar improper dihedrals $\chi_0 $ are also given the values adopted in the structure. The energy scale is set by  $\epsilon$, with  $\epsilon_r=100 \epsilon/\AA^2$, $\epsilon_\theta=80 \epsilon/{\rm rad}^2, \epsilon_\chi= 40 \epsilon/{\rm rad}^2$ for planar dihedrals, and $\epsilon_\chi= 10 \epsilon/{\rm rad}^2$ for other improper dihedrals.

The dihedral angle terms influence the local secondary structure of the protein. The values of $\phi_0$ are set to those in the experimentally-obtained structure. This model provides more stabilizing energy to backbone dihedrals, where the values of $\epsilon_{\rm bb}$ and $\epsilon_{\rm sc}$ are assigned values such that $\epsilon_{\rm bb}/\epsilon_{\rm sc}=2$. 

Non-bonded interactions  include stabilizing interatomic contacts, as well as steric repulsion between atom pairs that are not in contact in the native structure. In this simplest form of the model, non-bonded interactions are all described by isotropic pairwise interactions. Contact interactions are defined between any two atoms that are in contact in the native structure. Native contacts are defined based on the shadow algorithm.~\cite{noel2012shadow} In this algorithm, any two atoms are considered ``in contact'' in the experimental structure if they are separated by less than 6\,\AA, not connected by less than four bonds, and not occluded by any other atoms. All native contacts are given Lennard-Jones-like interactions, with minima set to $\sigma_{ij}$, the interatomic distances found in the experimentally-obtained structure. Consistent with theoretical analyses,~\cite{BRYNGELSON:1995p1708,EastwoodWolynes2001,Lammert:2011p11758}  $\epsilon_{\rm c}$ is defined such that the amount of energy in the contacts is twice the energy in the dihedrals:
\begin{equation}
\frac{\sum \epsilon_{\rm c}}{\sum \epsilon_{\rm bb} + \sum \epsilon_{\rm sc}}=2,
\end{equation}
and
\begin{equation}
\sum \epsilon_{\rm c} + \sum \epsilon_{\rm bb} + \sum \epsilon_{\rm sc}=N\epsilon .\end{equation} 
The inverse 12th power steric repulsion terms between all non-contact atoms have $\sigma_{\rm nc}=2.5\,\AA$ and $\epsilon_{\rm nc}=0.1\epsilon$.

\subsection{Protein folding thermodynamics}
For many protein sequences there is a well-defined folded state that is typically associated with a globular structure.  In this structure/state, there are many interactions  that contribute to the stability of the folded protein.  Interestingly, many proteins reversibly fold and unfold under constant temperature conditions. When the protein is folded, excluded volume interactions and topological constraints strongly limit the configurations that may be adopted, and the folded state may be described as a small ensemble (with low configurational entropy) of structurally-similar configurations. In contrast, the unfolded state is characterized by an extended peptide chain, where there is only a small number of native contacts that transiently form and break. As a result, the unfolded protein can explore a vast range of configurations. Accordingly, the folded state has low enthalpy and a low entropy, while the unfolded state has a higher enthalpy and higher entropy. 

The relative balance between folded and unfolded states depends on the thermodynamic stability of each state. In the canonical ensemble, systems are driven toward macrostates that minimize the free energy. We will consider the Helmholtz  free energy $F=E-TS$, where $E$ is the energy, $S$ is the entropy, and $T$ is the temperature.
At high temperatures, there is a greater  weight given to entropy, such that systems are driven toward high entropy states, thus lowering the free energy. At low temperatures, the influence of entropy is minimal, and  systems are driven toward low energy states. In the case of protein folding,  there exists a temperature for many proteins at which the folded and unfolded states have equal free energies, and are therefore equally probable. At this temperature, called the folding temperature $T_f$, the protein will reversibly interconvert between folded and unfolded states indefinitely.

There are several ways to identify the folding temperature of a protein. First, we can define a reaction coordinate (or order parameter) such as the fraction of native contacts  that distinguishes between the two states. We  simulate the system at many temperatures and determine the temperature at which the reaction coordinate values associated with the folded state occurs with the same probability as the values associated with the unfolded state.  Another way to identify $T_f$   is by the behavior of the specific heat at constant volume $C_V$. For temperatures less (greater) than $T_f$, the protein is predominantly in the folded (unfolded) state. For temperatures below or above $T_f$, the macrostate (folded or unfolded) does not depend strongly on the temperature, and the specific heat will be small. As the temperature approaches the folding temperature, the protein abruptly undergoes a pseudo-first-order phase transition between the folded and unfolded states. This rapid shift is accompanied by a very large change in the energy, which will manifest itself as a peak in the specific heat. The peak in $C_V$ indicates the pseudo-phase transition temperature, in this case the folding temperature. 

\begin{figure}[htbp!]
\centering
	\includegraphics[width=3.4in]{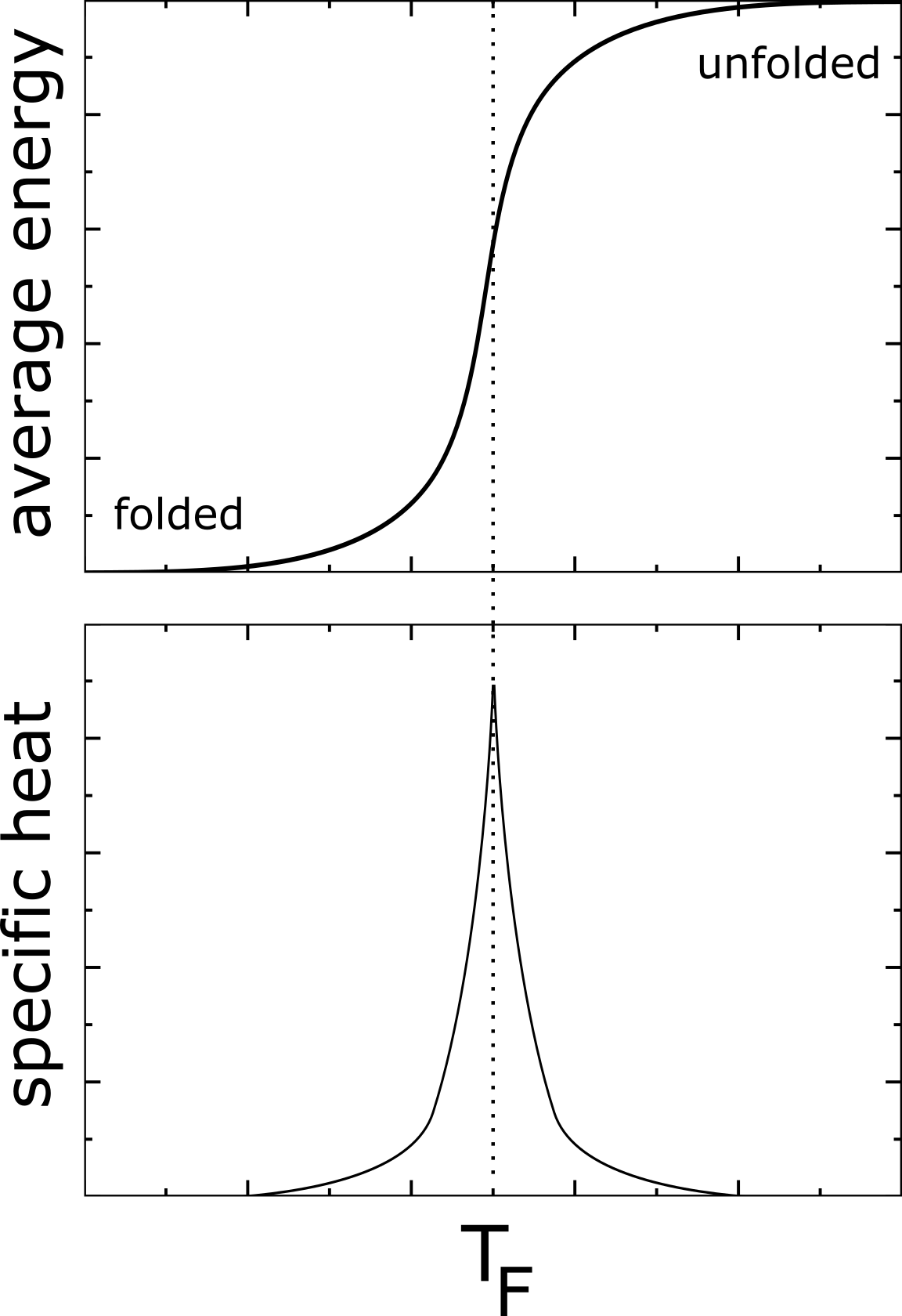}
\caption{Schematic of the temperature dependence of the average energy and specific heat  for a protein. The specific heat  exhibits a large peak around the folding temperature $T_f$.}
\label{ptn_thermo}
\end{figure}

\section{Diffusive motion and biomolecular rates \label{sec.:diffusion}}

To introduce  how the concept of diffusion is used to study the dynamics of protein folding, we  first discuss some  principles of diffusive dynamics. 

\subsection{Fundamentals of diffusive dynamics}

Diffusion is ubiquitous in  fields ranging from physics and chemistry, to economics and other social sciences. From a macroscopic view, we can treat diffusion in terms of the net movement of a quantity due to  a gradient in concentration, which drives particles to lower-concentration regions. Fick's first law formalizes the  relation between the diffusion coefficient  $D$, concentration $\rho$, and flux as~\cite{pathria2011statistics}
\begin{equation}
\mathbf{J} = -D \boldsymbol{\nabla} \rho, \label{eq:Ficks}
\end{equation}
where $\mathbf{J}$   is the diffusion flux vector.
By applying the principle of mass conservation in a closed system, and assuming a constant diffusion coefficient, we can derive the second Fick's law~\cite{pathria2011statistics}
\begin{equation}
  \frac{\partial \rho}{\partial t} = D \nabla^{2} \rho.
  \label{eq.:snd-fick}
\end{equation} 

Although Eqs.~\eqref{eq:Ficks}  and \eqref{eq.:snd-fick} provide a deterministic description of diffusion, the motion of an individual particle is largely random and controlled by noise. Historically, the erratic movement of pollen particles in a motionless water droplet first caught the attention of Robert Brown, and  the random motion that underlies diffusion is now known as Brownian motion.  Brownian motion was subsequently analyzed by Einstein in one of his \textit{annus mirabilis}' papers.~\cite{einstein1905motion, einstein-brownian} 
The relation between random thermal motion and    diffusion has been discussion by Einstein, Smoluchowski, Langevin, Fokker, among many others. 

For simplicity, we will  consider a one-dimensional representation of a diffusive system and write  the Brownian motion in terms of the Langevin equation as 
\begin{equation}
  m \frac{d^{2}x}{{dt}^{2}} = - \alpha \frac{dx}{dt} + F_{r}(t),
  \label{eq.:langevin1d}
\end{equation}
where $\alpha$ is the friction coefficient and $F_{r}(t)$ is a  random force. This random term arises due to  collisions with the environment and has the  properties that there is  no directional bias, $\langle F_{r}(t) \rangle = 0$, and sequential collisions are random in direction and amplitude, $\langle F_{r}(t) F_{r}(t^{'})\rangle = B\delta(t-t^{'})$. We divide both sides of Eq.~\eqref{eq.:langevin1d} by $m$ and rewrite it in terms of the particle velocity to find
\begin{equation}
  \frac{dv}{dt} = - \gamma v + \zeta(t),
  \label{eq.:langevin1d-v}
\end{equation}
where $\gamma = \alpha/m$ and $\zeta(t) = F_{r}(t) / m$. Equation~(\ref{eq.:langevin1d-v}) obeys the same conditions as Eq.~(\ref{eq.:langevin1d}), and thus  $\langle \zeta(t) \rangle = 0$ and $\langle \zeta(t) \zeta(t^{'})\rangle =  \Gamma \delta(t-t^{'})$, with $\Gamma = B / m^{2}$. By applying these conditions to the solution of Eq.~(\ref{eq.:langevin1d-v}), we find that the solution of  Eq.~(\ref{eq.:langevin1d-v})  is
\begin{equation}
  \langle v^{2} \rangle - \langle v \rangle^{2} = \frac{\Gamma}{2\gamma} \left( 1 - e^{-2 \gamma t} \right ),
  \label{eq.:sol-langevin1d-v}
\end{equation}
which at long times  approaches  $\langle v^{2} \rangle - \langle v \rangle^{2} = \Gamma/(2\gamma)$.
By applying the equipartition theorem to the first term, we obtain  the explicit temperature dependence
\begin{equation}
  \Gamma = \frac{2 \gamma k_{\rm B}T}{m},
  \label{eq.:gamma-langevin-T}
\end{equation}
and consequently $B = 2 \alpha k_{\rm B}T$.
By using the same strategy, we can determine the mean-square displacement of a particle that is undergoing Brownian motion:
\begin{equation}
  \langle x^{2} \rangle - \langle x \rangle^{2} = \frac{\Gamma}{\gamma^{2}} \left [t - \frac{2}{\gamma} \left( 1 - e^{-\gamma t} \right ) + \frac{1}{2\gamma} \left( 1 - e^{- 2 \gamma t} \right ) \right],
  \label{eq.:sol-langevin1d}
\end{equation}
where the left-hand side is the mean squared displacement as a function of time. In the long-time limit, the first term of Eq.~(\ref{eq.:sol-langevin1d}) is dominant and the mean-square displacement reduces to~\cite{tome2015stochastic, islam2004Einstein}
\begin{equation}
  \langle x^{2} \rangle - \langle x \rangle^{2} = 2Dt,
  \label{eq.:diff-brown}
\end{equation}
where $D = \Gamma/(2 \gamma^{2}) $ and  $D = B/(2 \alpha^{2})$.
From Eq.~(\ref{eq.:gamma-langevin-T}) we can write the relation between the diffusion coefficient and the temperature as
\begin{equation}
  D = \frac{k_{\rm B} T}{\alpha}.
  \label{eq.:diffusion-temp}
\end{equation}
Equation~(\ref{eq.:diffusion-temp}) is also applicable for two and three dimensions.
If we substitute the friction coefficient for a spherical particle of radius $r$ moving in a liquid of viscosity $\mu$, Eq.~(\ref{eq.:diffusion-temp}) becomes
\begin{equation}
  D = \frac{k_{\rm B} T}{6 \pi \mu r}.
  \label{eq.:diffusion-einstein}
\end{equation}

\subsection{Applying diffusive concepts to biomolecules \label{sec.:Kramers}}

Although the early studies of diffusion provided fundamental insights into the principles that govern stochastic processes, it was Hendrik A.\ Kramers who  extended these concepts to analyze the rates of chemical reactions.~\cite{Kramers:1940p2680,brinkman-1956,brinkman-II-1956} By building on Einstein's results, he    derived general diffusion equations that  describe the dynamics in the low and high viscosity regimes. The latter, overdamped regime is most relevant to the study of biomolecular dynamics and  will be used here to describe the dynamics of large-scale collective processes. 

Protein folding is a self-organizing process that is well described in terms of chemical reaction concepts.~\cite{Socci:1996p1697} It is useful to   categorize each accessible structure of a biomolecule as belonging to the unfolded, partially folded, or fully folded ensemble. We use the number of native contacts as the reaction coordinate, and an  ensemble of configurations is associated with a common number of contacts. If these ensembles are ordered in terms of their similarity to the folded state,  the number of conformations that can be accessed decreases dramatically as the folded state is approached.~\cite{BRYNGELSON:1989p1699,Leopold:1992p11568,BRYNGELSON:1995p1708} The distribution of configurations then resembles a funnel-like entity, where the global minimum corresponds to the folded structure. Based on the principle of minimal frustration,~\cite{BRYNGELSON:1989p1699,Leopold:1992p11568,BRYNGELSON:1995p1708} the energetic roughness on the funnel must be small relative to the difference in energy between the folded (the bottom of the funnel) and unfolded states (the top of the funnel). In this interpretation, the kinetics of folding  is  governed by the overall slope of the funnel, as well as its roughness, where the latter controls the diffusive properties of the molecule. 

Although protein energy landscapes have many dimensions,  it is often sufficient to consider the free energy as a function of only a small number of coordinates. 
These low-dimensional free energy landscapes  have two dominant minima that correspond to the unfolded and folded states,   similar to the descriptions of products and reactants when discussing chemical kinetics. Chemical kinetics are often described using single atomic distances or angles. In contrast, folding is a collective process that typically requires more elaborate metrics to quantify. A  widely used coordinate for describing folding is the fraction of native contacts that are formed as a function of time.~\cite{Clementi:2000p3234,whitford2009all,SMOG2} As the protein folds, more of the native (folded) contacts are formed, and the coordinate will adopt larger values. 
In many proteins,  this reaction coordinate   captures the basic diffusive properties of the folding process,~\cite{Cho:2006p9736} which allows  the system to be described in terms of the Fokker-Planck equation. This equation describes the time evolution of a stochastic process subject to a deterministic drift,~\cite{risken1996fokker}
\begin{equation}
  \frac{\partial}{\partial t} P(x, t) = \left[ - \frac{\partial}{\partial x}  v  + \frac{\partial^{2}}{{\partial x}^{2}} D  \right ] P(x, t),
  \label{eq.:fokker-planck}
\end{equation}
where $P(x, t)$ is the probability density of $x$, $v$ is the drift velocity (associated with the external force), and $D$ is the diffusion coefficient. 
Although  Eq.~\eqref{eq.:fokker-planck} is given in terms of the spatial coordinate $x$, this relation can be used to describe other stochastic processes, and   $x$ can represent either a spatial coordinate or a generalized reaction coordinate. 
In addition, the diffusion coefficient $D$ can depend on the value of $x$. 

For short time scales, the solution of Eq.~(\ref{eq.:fokker-planck}) is given by~\cite{risken1996fokker}
\begin{equation}
  P(x,t) = - \frac{1}{\sqrt{4 \pi D(x_{c})t}} \exp{\left[ - \frac{\left( x - x_{c} - v(x_{c})t \right)^{2}}{4D(x_{c})t} \right]}, 
  \label{eq.:fokker-planck-solution}
\end{equation}
for the initial condition $P(x,t=0) = \delta(x_{c})$. Equation~\eqref{eq.:fokker-planck-solution} represents  a Gaussian distribution, initially centered at $x_{c}$, moving with velocity $v(x_{c})$, where the width of the Gaussian $\sigma$ increases as the square root of $t$ ($\sigma(t) = \sqrt{2D(x_{c})t}$). 
The drift and diffusion coefficients can be expressed as
\begin{equation}
  v(x_{c}) = \frac{\langle x(t_{2}) \rangle - \langle x(t_{1}) \rangle}{\Delta t}
  \label{eq.:drift}
\end{equation}
and
\begin{equation}
  D(x_{c}) = \frac{\sigma^{2}(t_{2}) - \sigma^{2}(t_{1})}{2 \Delta t},
  \label{eq.:diffusion}
\end{equation}
where $\Delta t = t_{2} - t_{1}$. In principle, Eqs.~\eqref{eq.:drift} and \eqref{eq.:diffusion}  should be evaluated in the limit $\Delta t \rightarrow 0$ to obtain the drift and diffusion coefficients from a given data set. However, in real-world applications, $\Delta t$ only has to be small enough to ensure convergence of both quantities.

By using Eqs.~\eqref{eq.:drift} and \eqref{eq.:diffusion}, it is possible to numerically extract the diffusion and drift coefficients directly from a time series of values for a specific reaction coordinate $x(t)$.~\cite{Yang:2006p3338}
We will provide an example of how to use these relations to quantify diffusive dynamics, which we call the drift-diffusion (DrDiff) approach.~\cite{oliveira2018stochastic,freitas2019drift}$^{,}$ \footnote{The associated computational tools are available for download at \protect \url{https://github.com/ronaldolab/DrDiff} } To use this approach, one discretizes the reaction coordinate values into bins, where each bin is centered around  $x_{c}$ with a width of $\delta x_{c}$. By using these binned time values, time-dependent distributions are calculated over the interval $[t_{\rm initial}, t_{\rm final}]$. The functional form given by Eq.~(\ref{eq.:fokker-planck-solution}) is then fit to each distribution to provide an estimate of the position of the Gaussian center and standard deviation for each  time $\Delta t$. Linear regressions for $\sigma^{2}(t)$ and $x_{c}(t)$ are then evaluated using all the values obtained from the  time interval considered.

For systems that are  described well  in terms of diffusion on a one-dimensional landscape, we can obtain several  relations between diffusion, drift, free energies and rates.  For example, the free energy profile can be extracted from the drift velocity and diffusion coefficient using~\cite{Kopelevich2005}
\begin{equation}
  F(x)/k_{\rm B}T = - \!\int_{x_{\rm ref}}^{x} \frac{v(x')}{D(x')} dx' + \ln D(x) + \mbox{constant},
  \label{eq.:free-energy-stochastic}
\end{equation}
where the additive constant is related to the arbitrary free energy reference state $x_{\rm ref}$. Equation~\eqref{eq.:free-energy-stochastic} can be derived by assuming the equilibrium probability density $P_{\rm eq}$ is a solution of the steady-state Fokker-Planck equation, Eq.~\eqref{eq.:fokker-planck} and that $P_{\rm eq}(x) \propto \exp[-F(x) / k_{\rm B}T]$, where $F(x)$ is the free energy.~\cite{Kopelevich2005}  In addition, the mean first-passage time $\tau_{f}$ between two points on the profile, which is inversely related to the rate, is given by~\cite{szabo1980first}
\begin{equation}
  \tau_{f} = \!\int^{x_{\rm fold}}_{x_{\rm unf}} dx \int^{x}_{0} dx' \frac{e^{\beta \left [ F(x) - F(x') \right ]}}{D(x)},
  \label{eq.:first-passage-time}
\end{equation}
where $F(x)$ is the coordinate-dependent free energy profile, $D(x)$ is the coordinate-dependent diffusion coefficient, and $\beta = 1/k_{\rm B}T$. In this  context, it is assumed that the coordinate increases as a function of the reaction (folding) and that there is a lower bound of zero.

Equation~(\ref{eq.:first-passage-time}) can be used to obtain folding time scales if $x_{\rm unf}$ and $x_{\rm fold}$ define the unfolded and folded state minima on the free energy profile. The intervening barrier defines the transition state ensemble, which is the collection of configurations that have values  of the coordinate for which the free energy is maximal.  In the following, we will use the fraction of native contacts as the coordinate for folding, because it has been shown to exhibit diffusive properties and captures the rate-limiting barrier for many systems.~\cite{Cho:2006p9736}

\subsection{WHAM - Weighted Histogram Analysis Method}

The calculation of   the thermodynamic properties of a system can be reduced to determining the density of states.   We will employ the  Weighted Histogram Analysis Method (WHAM) to combine histograms from multiple simulations performed with different thermodynamic parameters (temperature). Instead of  providing a   derivation, we  provide  a brief introduction to  WHAM, so that  readers may   appreciate this statistic mechanics tool. 

The potential of mean force~\cite{kirkwood1935statistical, ROUX:1995p11834} is widely used as a measure of the free energy change during  biomolecular processes. In terms of the probability distribution $p(\xi)$, we may obtain the potential of mean force from the relation
\begin{equation}
  W(\xi) = W(\xi^{*}) - k_{\rm B}T \ln \left [ \frac{ p(\xi) }{ p(\xi^{*}) } \right ],
  \label{eq.:PMF}
\end{equation}
where $\xi^{*}$ and $W(\xi^{*})$ are arbitrary constants, $\xi$ is the system reaction coordinate, and $W$ is the potential of mean force. 
This approach is useful when there is sufficient Boltzmann sampling, although the accuracy of the calculation is limited by the quality of the  data. 
WHAM is a powerful technique that allows one to extract unbiased distributions from data sets that were obtained with or without biasing forces. As suggested by the name, WHAM estimates the relative weights of the sampled histograms to generate a set of unbiased free energy profiles. The algorithm applies the following arguments. For a set of $Z$ histograms, the average distribution function is given by~\cite{FERRENBERG:1988p1748, FERRENBERG:1989p1749, KUMAR:1992p1751}
\begin{equation}
  \left \langle p(\xi) \right \rangle = \frac{ \sum_{i=1}^{Z} n_{i} \left \langle p(\xi) \right \rangle_{i}}{ \sum_{j=1}^{Z} n_{j} e^{- \beta \left[ w_{j}(\xi) - f_{j} \right] } },
  \label{eq.:probWHAM}
\end{equation}
where $n_{i}$ is the number of  snapshots used to calculate a histogram/distribution from the $i$th simulation: $\left \langle p(\xi) \right \rangle_{i}$. From Eq.~(\ref{eq.:probWHAM}), the free energy of the $i$th simulation can be calculated from
\begin{equation}
  \beta f_{i}  = - \ln \!\int \left \langle p(\xi) \right \rangle e^{-\beta w_{i}(\xi)} d\xi.
  \label{eq.:free-energy-WHAM}
\end{equation}

Equations~(\ref{eq.:probWHAM}) and~(\ref{eq.:free-energy-WHAM}) were derived by minimizing the sampling errors by considering the overlapping regions of the probability distributions.~\cite{KUMAR:1992p1751} In the WHAM algorithm, Eqs.~\eqref{eq.:probWHAM} and~\eqref{eq.:free-energy-WHAM}  are iteratively evaluated until a self-consistent solution is obtained, at which point $ p(\xi)$ is approximated by $\left \langle p(\xi) \right \rangle$. This solution yields temperature-dependent free energy profiles  as well as the specific heat as a function of temperature. We apply a version of WHAM that is distributed with the SMOG2 software package;~\cite{SMOG2} details can be found in the SMOG2 documentation.

\section{Results}

We discuss an introductory example of how we can quantify diffusive aspects of protein folding using a simplified model.  First, we will discuss how to identify the folding temperature for a given protein and model.  We  then discuss the calculation of free energy barriers and the calculation of diffusion coefficients from folding trajectories. We   close with a brief discussion of convergence considerations when applying molecular simulations to study dynamics. For our discussion we will simulate and analyze the dynamics of the protein Chymotrypsin Inhibitor 2 (CI2; PDB code 2CI2).~\cite{McPhalen1987paper}  CI2 was chosen because it has been widely utilized as a model protein for the study of folding dynamics. In addition, it is a relatively small protein that can be easily simulated, making it an excellent system for demonstrating the techniques and ideas associated with the analysis of diffusion in proteins.

All simulation results presented here are available at the SMOG2\_tutorial repository. \footnote{\protect \url{https://github.com/smog-server/SMOG2_tutorial}} The repository also provides a file with instructions on how to replicate the simulated trajectories and analysis. Due to the stochastic properties of biomolecular dynamics, individual time traces are not exactly reproducible. However, the statistical properties should be consistent with our discussion. 

\subsection{Finding the folding temperature \label{subsec.:folding-temperature}}

\begin{figure}[htbp!]
  \centering
	\includegraphics[width=3.4in]{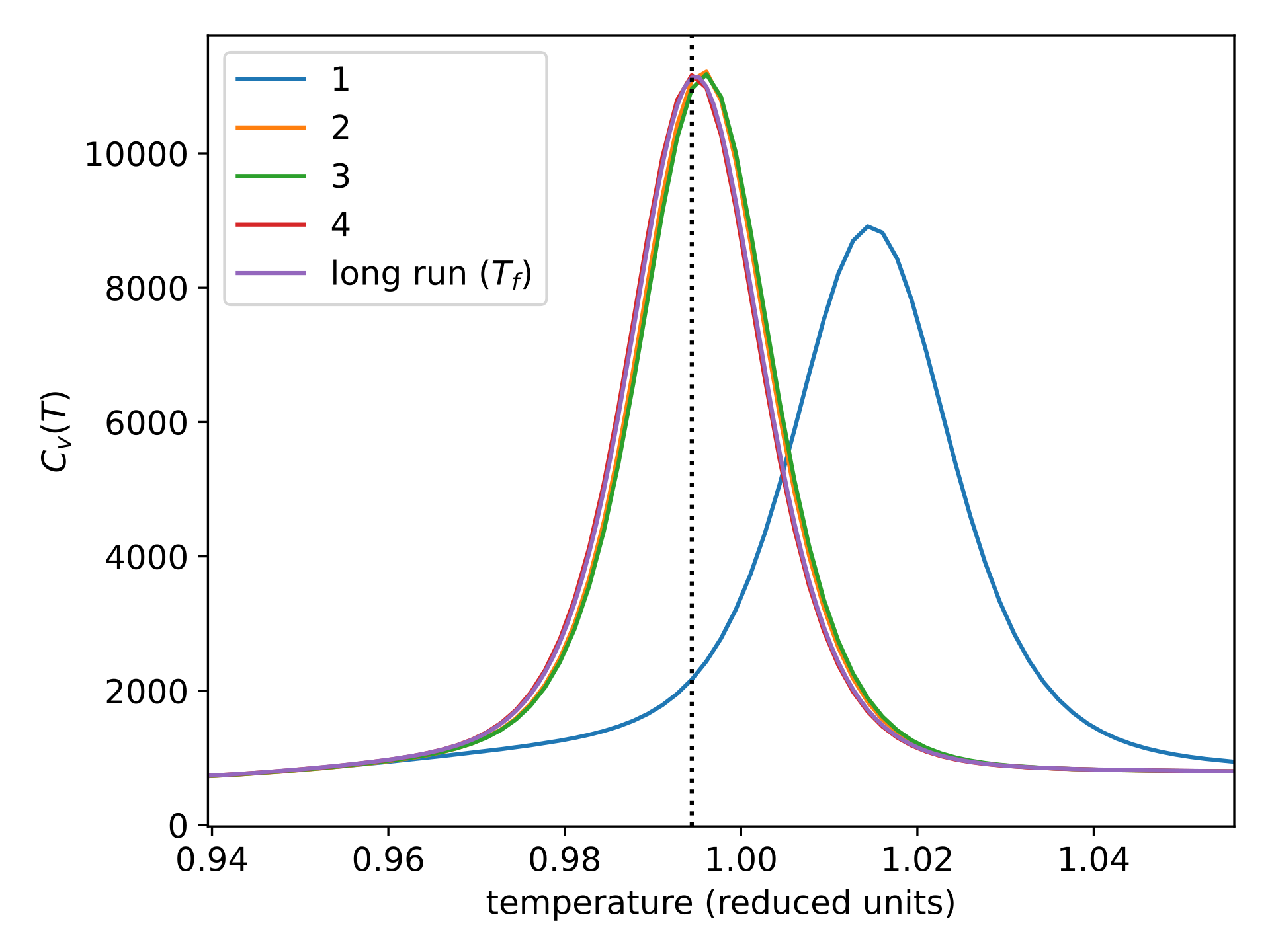}
  \caption{\label{fig.:specific_heat}
(Color online) The specific heat as a function of temperature (in reduced units)  for a structure-based model of the protein CI2. To find an initial estimate of the folding temperature, we   evaluate the temperature where $C_V$ reaches a peak value. As the sampling is enhanced (iterations 1--4), the peak maximum shifts and moves closer to the correct folding temperature value of $T_{f} = 0.994$.}
\end{figure}

To study protein folding/unfolding under  constant temperature conditions, it is necessary to first identify the folding temperature.  To do so, we typically start by performing several constant temperature simulations that span a wide range of temperatures. All simulations were  performed with the GROMACS software package.~\cite{Lindahl:2001, GROMACS2015}  GROMACS is a molecular dynamics simulation package that is widely used for the simulation of proteins, nucleic acids and lipids. This package integrates a given potential energy function as  in  Eq.~\eqref{eq.:SMOG-potential} to determine the time evolution of a system. To visualize the simulated trajectory, we  use  software such as VMD (Visual Molecular Dynamics). Because GROMACS does not allow for the use of reduced units (the Boltzmann constant is hard coded), a reduced temperature of 1 corresponds to a GROMACS temperature of 120. That is, in  reduced units, the Boltzmann constant is set equal to 1.  However, because Gromacs uses a value of $0.00831 \approx 1/120$ for the Boltzmann constant,  a numerical value of $\approx 120$ in Gromacs is equivalent to $k_BT=1$. In structure-based models,  room temperature corresponds to approximately 0.5 reduced energy units. For the first iteration (iteration 1 in Fig.~\ref{fig.:specific_heat}), the simulations were performed at 8 different temperatures ranging from 0.67 to 1.25 (in reduced temperature units).  WHAM was then used to combine the data from all temperatures and generate a specific heat curve (Fig.~\ref{fig.:specific_heat}; right-hand curve). This initial set of simulations gives  a pronounced peak  at $T\approx1.015$. Because it is possible that poor (non-Boltzmann) sampling can lead to artificial peaks in $C_V$, we next performed additional simulations near the candidate folding temperature.  In iteration 2,  six additional simulations were performed for temperatures  ranging from 0.99 to 1.03, and the simulations from both iterations were then combined using WHAM.  Perhaps surprisingly, we see a clear shift in the $C_V$ peak to lower temperatures. Based on this result, we performed five additional simulations (iteration 3) near the new candidate folding temperature  of $\approx 1.0$. In this example, we repeated the process for a total of four iterations. Between the third and fourth iterations there were minimal changes in  $C_V$, which implies that $T_{f} \approx 0.994$. 

\begin{figure}[htbp!]
  \centering
	\includegraphics[width=3.4in]{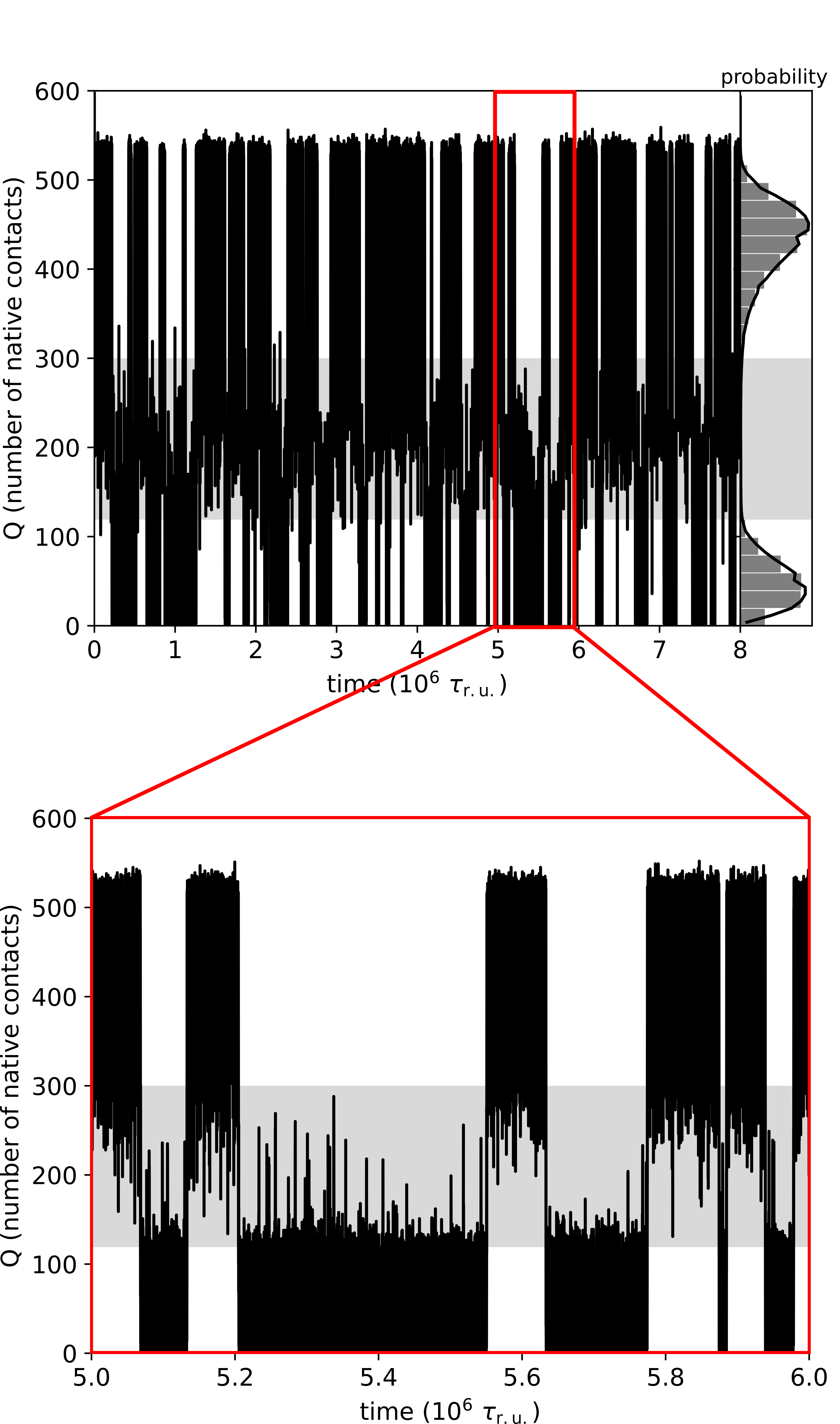}
  \caption{\label{fig.:trajectory}
  Number of native contacts $Q$ as a function of time (top). In this trajectory, there are  $\approx 80$ transitions between the folded (high $Q$) and unfolded (low $Q$) states. The inset on the right shows the corresponding probability density.
  The bottom plot shows a zoomed-in perspective, which highlights the transition events and relative stabilities of the folded and unfolded states.
  The transition state ensemble in the gray region corresponds to configurations  that successfully cross the underlying free energy barrier.}
\end{figure}

After identifying the folding temperature, we typically  perform a much longer simulation at the folding temperature to obtain a large number of spontaneous folding and unfolding events. We performed a single simulation at the folding temperature of $10^{9}$ time steps, which is 100 times longer in duration than the initial simulations.  We will refer to this long simulation as the ``trajectory.'' Figure~\ref{fig.:trajectory} shows the number of native contacts  $Q(t)$ as a function of time. A zoomed-in view  of the simulation indicates there is a clear separation between the folded (high $Q$) and unfolded (low $Q$) states where   abrupt transitions  occur. In total,  $\approx 80$ transitions between folded and unfolded states were found. The plot on the top right  in Fig.~\ref{fig.:trajectory} shows two well defined peaks in the probability distribution. The peak at $Q \approx 450$ is associated with the folded state, and the one centered at $Q = 50$ corresponds to the unfolded state. For reference, the native structure has 597 native contacts. The fact that the ensemble of folded configurations only has $\approx 80$\% of the native contacts formed may be surprising. This difference between the theoretical maximum and the most probable value can be understood in terms of simple thermodynamic considerations. That is, when all contacts are formed the protein adopts a highly compact form, which is associated with strong steric interactions that confine the chain.  In other words, forming all contacts simultaneously is entropically disfavored. 

\subsection{Calculating drift velocities and diffusion coefficients}

Unlike applications of diffusion  that describe the flux of particles that arise from a concentration gradient, diffusion in protein folding describes the probability flux between the members of a  conformational ensemble. To describe this ensemble, it is often suitable to choose    reaction coordinates to measure the ``folded-ness'' of the system. 
Although there is no guarantee that a given coordinate will be suitable for the analysis of diffusion, we will find that the number of contacts is suitable for describing the kinetics and thermodynamics of this protein. 

Although textbooks often describe a diffusion coefficient as a constant, it can also be time and coordinate dependent.~\cite{Chahine:2007p1669, Oliveira:2010p10996, Hummer-PNAS-2010,freitas2019drift} In the context of protein folding, there have been many techniques applied to quantify diffusive properties~\cite{Schulten1981, Gruebele1999}  from experiments~\cite{Munoz1999, Kubelka:2004p10124} and simulations.~\cite{hummer2004transition, Yang:2006p3338, Krivov:2008p9677} To introduce  the analysis and interpretation of diffusive dynamics, we  apply the DrDiff approach described in Sec.~\ref{sec.:Kramers} to evaluate both the drift and diffusion coefficients, as well as the free energy profile from a simulated data set.  

\begin{figure}[htbp!]
  \centering
	\includegraphics[width=3.4in]{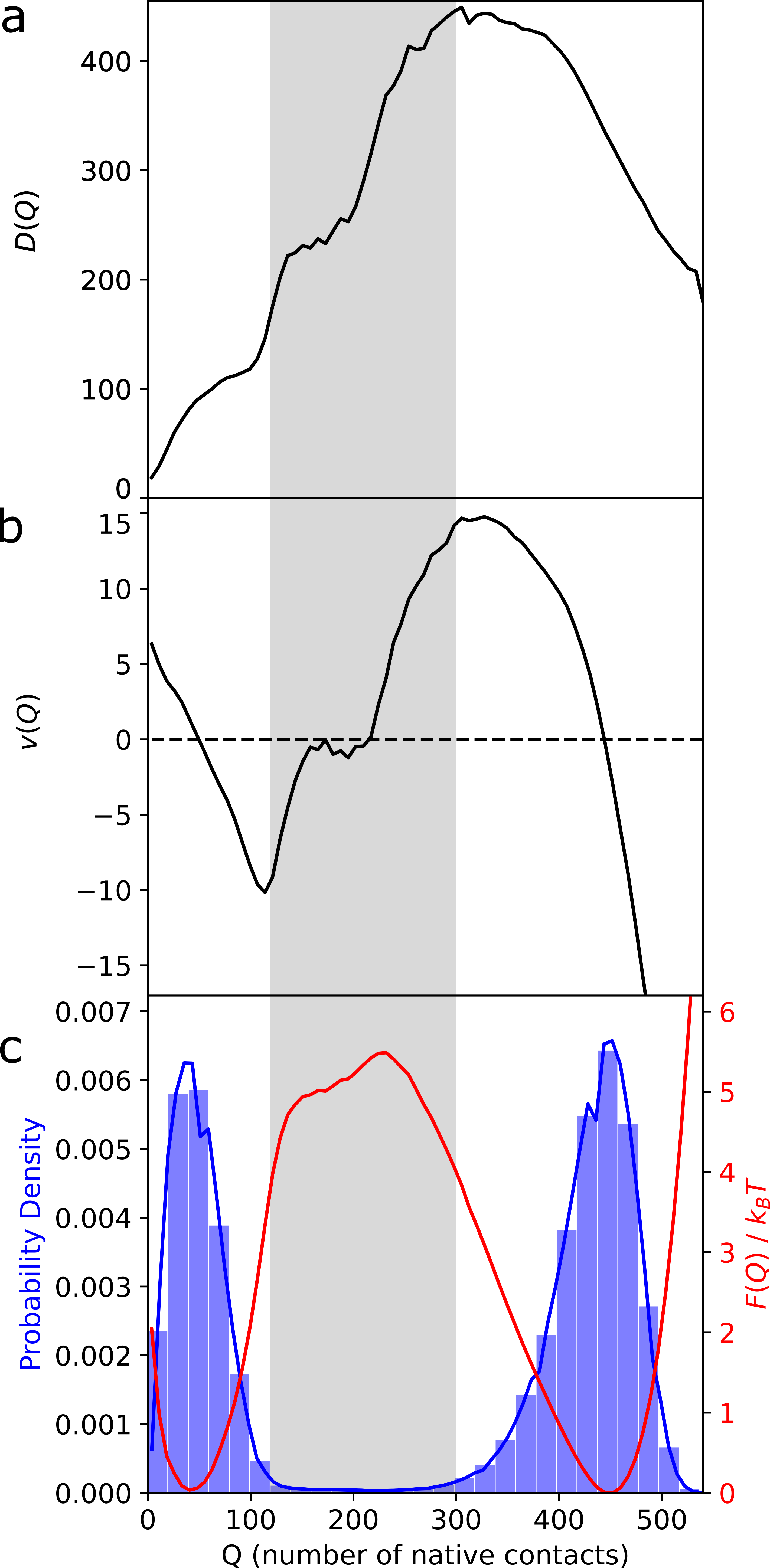}
  \caption{\label{fig.:free-diffusion-drift}
(a) Coordinate-dependent diffusion coefficient obtained using  DrDiff, Eq.~(\ref{eq.:diffusion}). The diffusion coefficient increases from the unfolded state until the end of the transition state (gray shaded area), and then decreases as the protein approaches the folded state.
(b) Coordinate-dependent drift coefficient extracted from the trajectory data. 
  The gray shaded area  shows the transition state, defined as the region where the free energy is within $k_{\rm B}T$ of its peak value at the barrier. 
  (c) Probability density as a function of the number of native contacts (blue, peaks near $Q=50$ and 450), alongside the free energy profile integrated from Eq.~(\ref{eq.:free-energy-stochastic}) (red, peak near $Q=250$).}
\end{figure}

We analyze a long constant-temperature simulation (see Fig.~\ref{fig.:free-diffusion-drift}) to estimate the coordinate dependent diffusion and drift coefficients. As mentioned in Sec.~\ref{subsec.:folding-temperature}, this simulation includes more than 80~folding/unfolding events where the protein spontaneously samples fully folded and unfolded configurations. Figure~\ref{fig.:free-diffusion-drift}(a) shows the diffusion coefficient calculated using Eq.~(\ref{eq.:diffusion}). The diffusion coefficient increases with the number of native contacts, and then decreases as the folded state is reached. 
We additionally calculate the drift velocity [see Fig.~\ref{fig.:free-diffusion-drift}(b)], which is zero    at  distinct values of $Q$. Intuitively, these states should correspond to free energy minima and maxima. The diffusion and drift coefficients were then used to estimate the free-energy profile using Eq.~(\ref{eq.:free-energy-stochastic}). Based on the time traces, there are two deep free energy minima that are separated by a clear barrier [Fig.~\ref{fig.:free-diffusion-drift}(c)]. 
By comparing the probability density peaks with the free energy curve, it is clear that the two minima are more highly sampled than the transition state, which  is the region where the free energy is no less than $k_{\rm B}T$ below the  peak value.

The increase of the diffusion coefficient from the unfolded state to the transition state [Fig.~\ref{fig.:free-diffusion-drift}(a)] reveals how entropic factors can influence diffusive properties.  That is, during folding, a protein must initially collapse, and it  becomes increasingly difficult to form additional contacts due to a high level of residual disorder in the chain. However, once a critical set of contacts is formed, the protein is sufficiently confined that the rearrangements that result in  formation of additional contacts are essentially the only allowable motions.

\subsection{Calculating the free energy barrier \label{subsec.:free-energy-barrier}}

\begin{figure}[htbp!]
  \centering
	\includegraphics[width=3.4in]{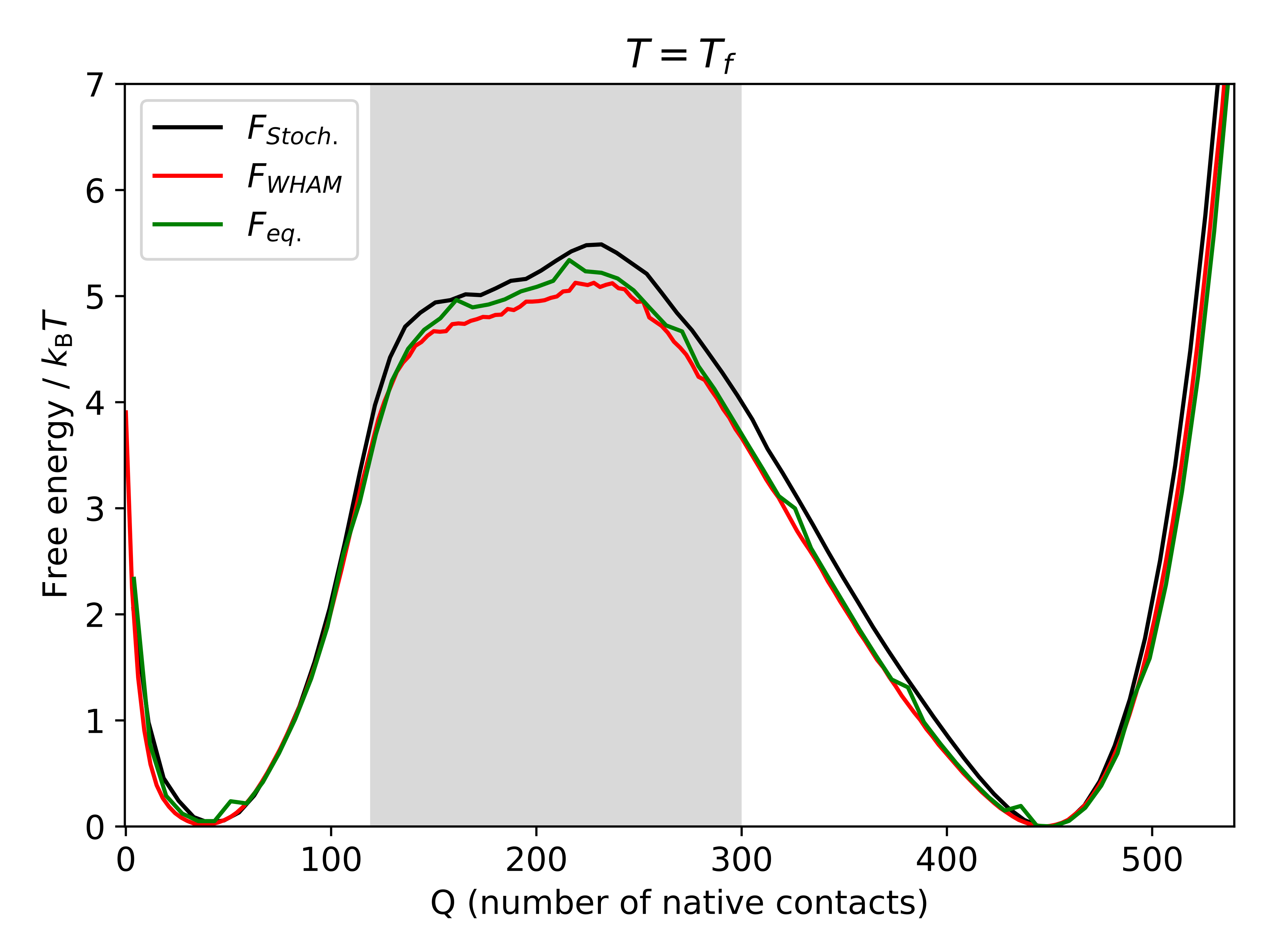}
\caption{\label{fig.:free-energy}
  (Color online) Comparison of the free energy profiles obtained by  three different techniques. The green curve presents the free energy obtained  from the trajectory probability density, using Eq.~(\ref{eq.:free-energy-equilibrium}). In black is the free energy integrated from the coordinate-dependent drift and diffusion coefficients using Eq.~(\ref{eq.:free-energy-stochastic}). Both are compared with the result from WHAM (red).
  The gray shaded area is the transition state ensemble.}
\end{figure}

There are several techniques to determine the free energy from a simulation. The most direct method is to evaluate the potential of mean force from the probability density, using
\begin{equation}
F_{\rm eq} = - \ln \left[ P(Q)  \right] + \mathbf{C},
\label{eq.:free-energy-equilibrium}
\end{equation}
where $P(Q)$ is the probability density for the chosen reaction coordinate $Q$, and $\mathbf{C}$ is an arbitrary constant. 
The green curve in  Fig.~\ref{fig.:free-energy} represents the free energy calculated from  Eq.~(\ref{eq.:free-energy-equilibrium}). 
Figure~\ref{fig.:free-energy} also displays the free energy profile calculated from multiple simulations at different temperatures that are combined using WHAM. In addition, the free energy was estimated based on Eq.~(\ref{eq.:free-energy-stochastic}), where the drift and diffusion coefficients were obtained from the DrDiff approach.  The three techniques were applied to demonstrate the suitability of the reaction coordinate for describing the diffusive process.

We find that the main features of the free energy profile at $T_{f}$ are the same using the three  techniques we considered. The positions of the minima and barrier are the same, although there are small   differences in the barrier heights. The   agreement suggests that the kinetics are well described in terms of one-dimensional diffusion along our chosen  reaction coordinate. We stress that agreement between the methods is not guaranteed for an arbitrary dynamical system. It is possible that some systems when described by certain coordinates may appear to exhibit sub-diffusive dynamics, even if the underlying dynamics is not truly subdiffusive.~\cite{krivov2010protein}

\subsection{Convergence and sampling considerations \label{subsec.:convergence}}

We now  discuss  the importance of both  duration and frequency of data storage/analysis  when estimating kinetic and thermodynamic quantities from simulations.  Because these considerations are not unique to a particular form of analysis, the general strategies can be applied to other areas as well.

\begin{figure}[htbp!]
  \centering
	\includegraphics[width=3.4in]{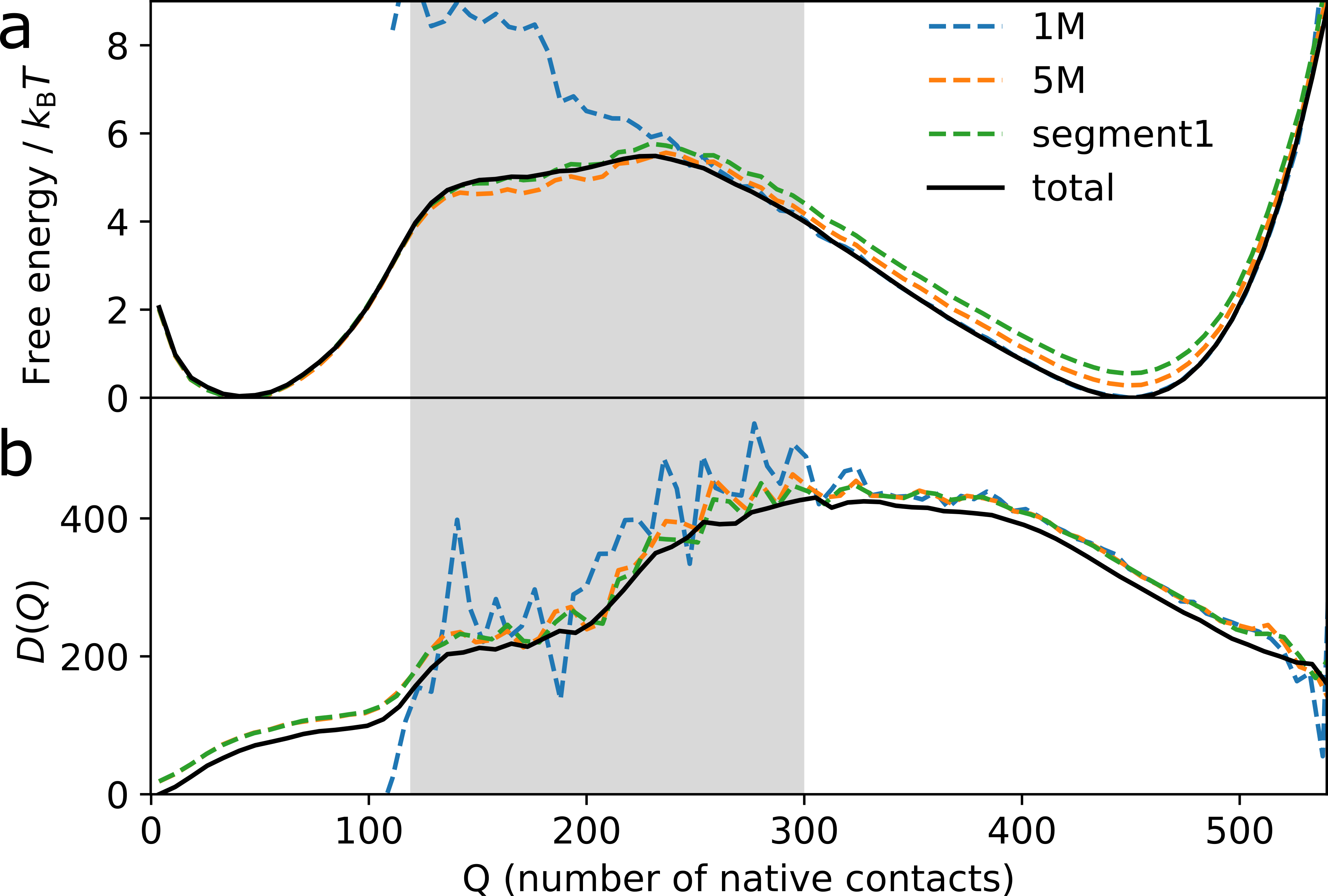}
  \caption{\label{fig.:free-diffusion-comparison-start}
  (Color online) Results from a complete trajectory at the folding temperature ($\sf total$) compared with results from three subsets. 
  (a) Free energy integrated from Eq.~(\ref{eq.:free-energy-stochastic}) for different trajectory segments (dashed lines) alongside the full trajectory (solid black line). The numbers are related to the file numbers available in the repository.
  (b) Corresponding coordinate-dependent diffusion coefficients obtained using  DrDiff  via Eq. ~\eqref{eq.:diffusion}.
  With no transitions, the $\sf 1M$ trajectory segment (first $10^6$ time steps) shows discrepancies in the transition state and low $Q$ values. All other analyzed segments show free energy profiles and diffusion coefficient curves that agree with those generated from analysis of the complete trajectory.
  The gray shaded area is the transition state region.}
\end{figure}

To illustrate a simple method for verifying convergence, we  analyze fragments of the long trajectory used for the previous calculations.  First, we analyzed the initial  $10^{6}$ time steps of the simulation using  DrDiff. The results will be referred to as $\sf 1M$, and  contain the first ten percent of the trajectory (file \textsf{Q-119.5.segment1.dat} in the tutorial repository). Figure~\ref{fig.:free-diffusion-comparison-start}(a) shows the free energy profile (dashed blue line) estimated from this data set. As expected, the free energy is undefined for small values of $Q$ because no unfolding transitions occurred in these initial frames, and hence the unfolded conformations were not represented. The diffusion coefficient  is also very noisy  in the transition state region and no values are available for the unfolded ensemble. Next, the first $5 \times 10^{6}$ trajectory steps, referred to as $\sf 5M$, were analyzed and the results are displayed in Fig.~\ref{fig.:free-diffusion-comparison-start} (dashed orange lines). This part of the simulation included only four folding/unfolding transitions. Nonetheless, the diffusion coefficients and free energies  from the $\sf 5M$ data set are comparable with those obtained from the complete trajectory (80~transitions). 

\begin{figure}[htbp!]
  \centering
	\includegraphics[width=3.4in]{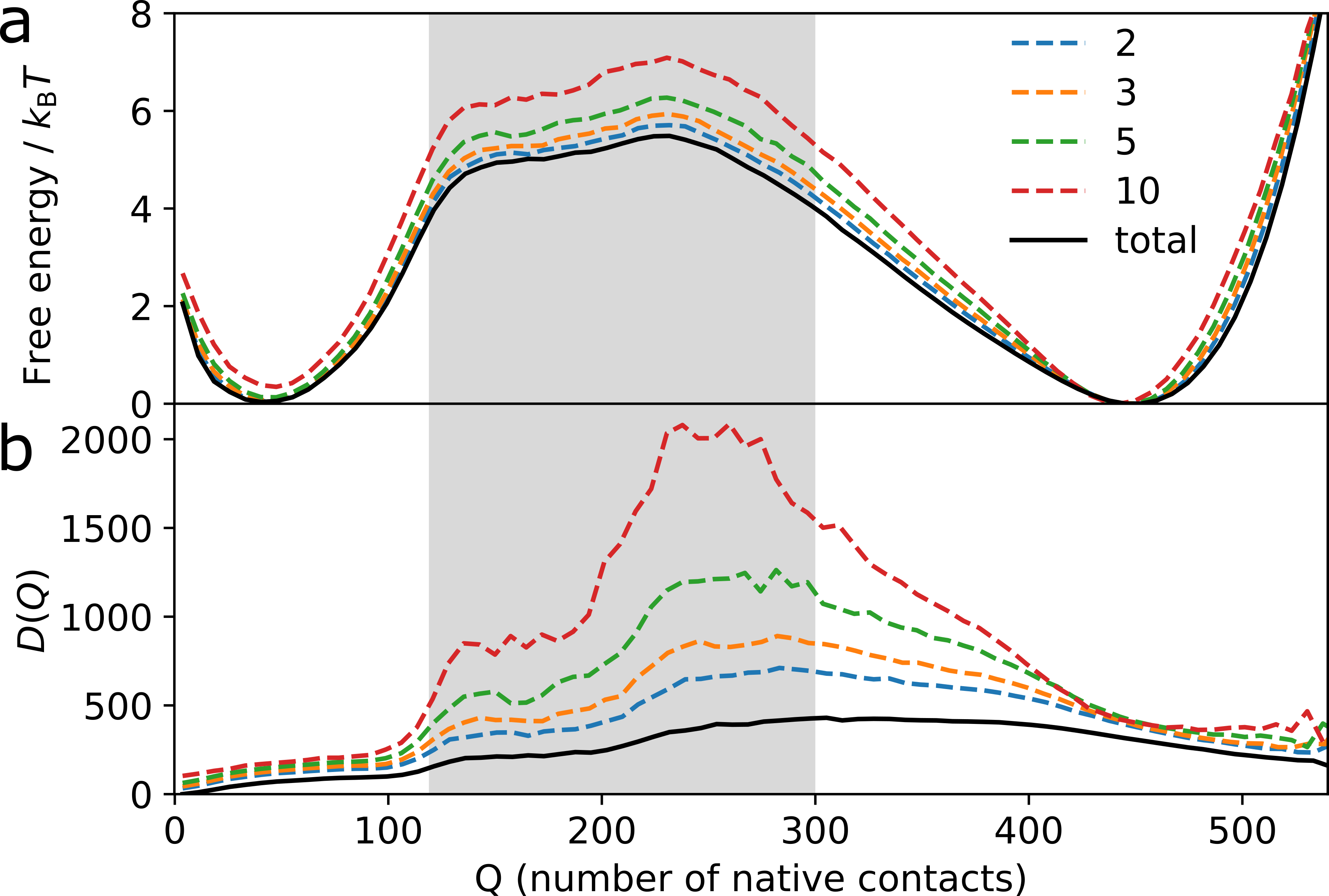}
\caption{\label{fig.:free-diffusion-comparison-stride}
  Results from a complete trajectory ($\sf total$) compared with the results of trajectory subsets saved after skipping every $w$ time steps, with $w$ varying from $2$ to $10$. 
  (a) Free energy integrated from Eq.~(\ref{eq.:free-energy-stochastic}) for five   values of $w$ applied to the trajectory at $T_{f}$. The positions of the free energy minima are the same, despite the different  values of $w$, while the barrier is reduced by  $\approx 2 k_{\rm B}T$ when $w$  is increased.
  (b) The corresponding coordinate-dependent diffusion coefficient is obtained via Eq.~(\ref{eq.:diffusion}). As $w$  increases, the deviations from the  total trajectory (black) increase because there are fewer sampled $Q$ values and because the  drift-diffusion analysis is based on short-time dynamics, some of which is missed as $w$ is increased. 
  The gray shaded area is the transition state region.}
\end{figure}

Another challenge when calculating diffusive properties is the choice of the proper frequency  of saving/analyzing simulated frames. For the DrDiff approach, it is necessary that the  reaction coordinate change by  small increments between saved configurations. To illustrate the influence of this point, results from the complete saved trajectory were compared to the results obtained when only every $w$th  frame is considered, where $w$ is known as the stride value.  A stride value of $w= 1$ means that all configurations were saved,  keeping the original trajectory intact. A stride value of $w=10$ means that every 10th configuration is saved, resulting in a trajectory 10 times smaller  with a time step 10 times longer.   Figure~\ref{fig.:free-diffusion-comparison-stride} shows how the different stride values impact the convergence of the diffusion coefficient values. As expected, larger deviations are observed as $w$  is increased.  Interestingly, the positions of the free energy minima and barrier were insensitive to the stride value [Fig.~\ref{fig.:free-diffusion-comparison-stride}(a)]. In contrast, the free energy barrier height changed by nearly $2k_{\rm B}T$ as $w$   was varied.  This dependence on $w$ highlights how, even when describing the same system, various thermodynamic and kinetic properties may converge  at different rates. Accordingly, it is always necessary  to verify the convergence and robustness of each  quantity analyzed. In  research studies, a more extended and iterative process than the one we discussed here is typically required to ensure that there is a satisfactory level of convergence.

\section{remarks}
There are many opportunities for physicists to identify novel problems in the biological sciences.  We have discussed how a spherical-cow-like model can be used to obtain insights into the diffusive dynamics of protein folding.  Our discussion  represents only the beginning of what may be explored with these models. Although    protein folding is a  mature field, the ideas developed in this context can be applied to a broad range of biological processes.  For example, the study of reaction coordinates and diffusion are providing insights into the relation between structure, energetics and dynamics in very large molecular assemblies, such as bacterial~\cite{LeviMethods} and eukaryotic~\cite{Freitas2021} ribosomes. We expect that as the physics community continues to investigate areas of biology, new questions will be posed and answered, which will reveal the organizing principles of molecular biological processes.

\begin{acknowledgments}
PCW was supported by NSF grant MCB-1915843. Work at the Center for Theoretical Biological Physics was also supported by the NSF (Grant PHY-2019745). FCF was financed by the Coodena\c{c}\~{a}o de Aperfei\c{c}oamento de Pessoal de N\'{\i}vel Superior - Brasil (Capes) - Finance Code 001. Financial support for RJO was provided by Funda\c{c}\~{a}o de Amparo \`{a} Pesquisa do Estado de Minas Gerais (FAPEMIG, APQ-00941-14) and Conselho Nacional de Desenvolvimento Cient\'{\i}fico e Tecnol\'{o}gico (CNPq, 438316/2018-5 and 312328/2019-2). 

\end{acknowledgments}


 \appendix*
  
\section{Simulation Visualization Tutorial}
Visual Molecular Dynamics (VMD) is a powerful three-dimensional (3D) molecular visualization and analysis software package~\cite{humphrey1996vmd}  that stands out in its capability to visualize   extremely large biomolecules and arbitrary graphics objects along with its long molecular dynamics trajectories. In addition, VMD includes a powerful scripting interface that supports Tcl-based commands~\cite{humphrey1996vmd,Ousterhout1994} for manipulating visualizations and performing analyses.

Platform-specific VMD can be downloaded \footnote{Available at \protect \url{https://www.ks.uiuc.edu/Research/vmd/}.} Once VMD is installed, it starts with three default  windows, VMD Main, VMD OpenGL Display, and VMD console. We explore the basic capabilities of VMD by going through  the steps required to generate various representations of the chymotrypsin inhibitor 2 protein.
\begin{enumerate}
\item Download the PDB formatted file of the chymotrypsin inhibitor 2 protein  (ID:2CI2) from the RCSB Protein Data Bank.~\footnote{The PDB - Protein Data Bank is located at \protect \url{https://www.rcsb.org/}.} This protein can be found using the search option on the PDB webserver. 

\item To load the atomic structure file (.pdb) of the molecule,  select File $\longrightarrow$ New Molecule in the VMD Main window and use the Browse option to choose the .pdb file in the Molecule File Browser window; then press Load. You will be able to see the 2CI2 molecule in the OpenGL Display, and the .pdb file will be listed in the Main window.
\item To modify the 3D visualization of the molecule, there are three modes  by which the mouse may be used to alter the perspective/view: Rotation, Translation and scaling. The mouse mode can be chosen from VMD Main $\longrightarrow$ Mouse $\longrightarrow$ R,T or S.
\item VMD also provides options to choose the mode of depth perception when viewing the molecule. VMD Main $\longrightarrow$ Display $\longrightarrow$ Perspective (strong depth perception) / Orthographic (low depth perception). We find that larger molecules are much easier to view with Orthographic, though Perspective is usually sufficient for small systems.

Although the default background color is black, it is often desirable to use other colors such as white. To change the background color in the OpenGL Display to white, go to VMD Main $\longrightarrow$ Graphics $\longrightarrow$ Colors $\longrightarrow$ Display (under Categories) $\longrightarrow$ Background (under Names) $\longrightarrow$ white.
\item VMD has many options available for graphical representations of the molecule and atom selections. The Graphical Representations window can be accessed from VMD Main $\longrightarrow$ Graphics $\longrightarrow$ Representations. This will provide options for Atom Selection, Drawing Method, Coloring Method, etc. The default representation is ``Lines'' for Style, ``Name'' for Color and ``all'' for Selection. It is possible to  either edit the default representation or create additional representations using Create Rep option.
\begin{itemize}
    \item The different options available for Drawing Methods can be found under Draw style. We have found that the most useful representations are VDW (sphere for each atom), Tube (representing only backbone traces) and  NewCartoon [protein secondary structure; similar to Fig.~1(b)].
    \item Similarly there are several options available to color the molecule or the selection based on for example, Name (atom type), ResType (residue type), Secondary Structure, and Backbone.
    \item The selection tab provides the keywords or single words you can string together using Boolean operators (e.g., not protein) to generate atom selections of your interest.
\end{itemize}
\item VMD also supports loading trajectories from molecular simulations into the loaded molecule. To load a trajectory of a molecule, first select the molecule in VMD Main window, followed by File $\longrightarrow$ Load Data into Molecule. The Molecule File Browser window pops up and Load Files for option will have the molecule selected. At this point, use the Browse option to select the trajectory file and press load. When following the accompanying tutorial, \footnotemark[1]  the trajectory files will have the suffix .xtc.  The loaded trajectory can then be viewed using the animation tools at the bottom of the Main window.
\end{enumerate}




\end{document}